\documentclass[11pt,a4paper]{article}

 \newcommand{\be}{\begin{equation}}
 \newcommand{\ee}{\end{equation}}
 \newcommand{\bl}{\begin{equation}\begin{array}{ll}}
 \newcommand{\el}{\end{array}\end{equation}}
 \newcommand{\bll}{\begin{equation}\begin{array}{lll}}
 \newcommand{\bdm}{\begin{displaymath}}
 \newcommand{\edm}{\end{displaymath}}
 \def\bea{\begin{eqnarray}}
 \def\eea{\end{eqnarray}}
 \def\barr{\begin{array}}
 \def\earr{\end{array}}
\def\p{\partial}
\def\d{\partial}

\def\half{\frac{1}{2}}
\def\quart{\frac{1}{4}}

\def\third{\frac{1}{3}}
\def\2third{\frac{2}{3}}
\def\4third{\frac{4}{3}}
\def\3quart{\frac{3}{4}}

\def\sixth{\frac{1}{6}}

\def\lim{\rightarrow}

\def\ba{\bar{a}}

\def\bV{\bar{V}}
\def\bk{\bar{k}}

\def\cL{{\cal L}}

\def\drho{\dot{\rho}}

\def\dsig{\dot{\sigma}}
\def\dpsi{\dot{\psi}}
\def\dal{\dot{\alpha}}
\def\dbe{\dot{\beta}}
\def\dga{\dot{\gamma}}
\def\dpsi{\dot{\psi}}
\def\dA{\dot{A}}

\def\df{\dot{f}}

\def\Lag{{\cal L}}
\def\Hag{{\cal H}}

\begin{document}
\raggedbottom

\title{{\bf On the Weyl - Eddington - Einstein affine gravity \\
 in the context of modern cosmology
 }}

\author{A.T.~Filippov \thanks{Alexandre.Filippov@jinr.ru}~ \\
{\small \it {$^+$ Joint Institute for Nuclear Research, Dubna, Moscow
Region RU-141980} }}

\maketitle

\begin{abstract}
 We propose new models of an `affine' theory of gravity in
 $D$-dimensional space-times  with symmetric connections.
 They are based on ideas of Weyl, Eddington and Einstein
 and, in particular, on Einstein's proposal
 to specify the space - time geometry by use of the Hamilton
 principle. More specifically, the connection coefficients are
 derived by varying a `geometric' Lagrangian that is
    supposed to be an arbitrary
 function of  the  generalized (non-symmetric)  Ricci curvature
 tensor (and, possibly, of other  fundamental  tensors) expressed in
 terms of the connection coefficients regarded as independent
 variables. In addition to the standard Einstein gravity, such a
 theory predicts dark energy (the cosmological constant, in the first
 approximation), a neutral massive (or, tachyonic) vector field, and
 massive (or, tachyonic) scalar fields. These fields couple only to
 gravity  and may generate dark matter and/or inflation. The masses
 (real or imaginary) have geometric origin and one cannot avoid their
 appearance in any concrete model. Further details of the theory -
 such as the nature of the vector and scalar fields that can describe
 massive particles, tachyons, or even `phantoms' -  depend on the
  concrete  choice of the geometric Lagrangian.
  In `natural' geometric theories,
 which are discussed here, dark energy is also unavoidable. Main
 parameters - mass, cosmological constant, possible dimensionless
 constants - cannot be predicted, but, in the framework of modern
`multiverse'  ideology, this is rather a virtue than a drawback of
 the theory. To better understand possible applications of the theory
 we discuss some further extensions of the affine models and analyze
 in more detail approximate (`physical') Lagrangians that can
 be applied to cosmology of the early Universe.
 \end{abstract}

\section{Introduction. Geometry and dynamics of the affine theory}
 In 1918-1923 H.Weyl \cite{Weyl}, A.Eddington \cite{Ed1}, \cite{Ed},
 and A.Einstein \cite{Einstein1}, \cite{Einstein2} proposed some
 generalizations
 of the Einstein gravity theory, which they considered as a `unified'
 affine field theory of gravity and electromagnetism. The final
 formulation was given by Einstein in three beautiful and concise papers
 \cite{Einstein1}, later summarized by Eddington \cite{Ed1} and
 Einstein \cite{Einstein2} and soon forgotten (see however brief
 discussions  in \cite{Erwin}, \cite{Pauli}). Recently, the simplest
 Einstein - Weyl model was reinterpreted in \cite{ATF}, where one can
 find a brief summary of the ideas and results of these papers that are
 of interest for today's concerns. Here we start with formulating two main
 models {\bf predicting} (in 1923!) the cosmological constant and
 a  {\bf massive} (or, {\bf tachyonic}) vector particle (`vecton')
 accompanying graviton
 This particle may be interpreted as a sort of dark matter (or,
 alternatively, as an {\bf inflaton})
  interacting with gravity only\footnote{
 If the vector field vanishes, both models reduce to the Einstein theory with
 nonvanishing cosmological constant.}.

 The great authors of these models did not study
 their beautiful new theory in detail and thus our first goal in
 \cite{ATF} was to look for simplest spherically symmetric solutions and,
 especially, to cosmological ones. Here we propose a generalization of
 the Einstein models to any dimension and argue that by its
 simple dimensional reduction we can also get
 massive (tachyonic) scalar mesons.
 We note that the second
 Einstein model (and its generalizations containing scalar inflatons)
 may be considered as the first approximation derived by expanding the
 square - root Lagrangian in powers of the vector field (see below).
 In this approximation we derive the general equations describing
 cosmologies and propose further approximation for which it is
 possible to find some exact analytic solutions (Section 3).
    Finally, we emphasize that
 Einstein's approach is based on a clearly formulated volition giving
 a beautiful theory. In the end of the paper
 we briefly discuss possible
 extensions of his ideas having relation to modern
 cosmology but first we consider his original approach and
 discuss its immediate and most economic generalizations.

 \bigskip
 The main idea of Weyl was that a unification of gravity with
 electromagnetism requires using a non-Riemannian
 {\bf symmetric connection}.
 He considered a special case of the connection depending on a
 symmetric tensor $g_{ij}$, which he identified with the Riemann
 metric, and on a vector $a_i$, which he tried to identify with the
 electromagnetic potential. It belongs to the simplest class
 suggested in \cite{ATF}:
 \be
 \label{a3}
 \gamma^m_{kl} = \half [g^{mn} (g_{nk,l} + g_{ln,k} - g_{kl,n}) +
 \alpha ( \delta^m_k \, a_l +  \delta^m_l \, a_k) -
   (\alpha - 2\beta) g_{kl}\, a^m ] \,,
 \ee
 where the commas denote differentiations, $g_{ij} g^{jk} =
 \delta^i_j$  and $a_m = g_{mk} a^k$.
 The Weyl connection corresponds to $\alpha = 1$, $\beta = 0$,
 for the Einstein connection $\alpha = 1/3$, $\beta = -1/3$.

 The curvature tensor can be defined
 without using any metric:
 \be
 \label{1}
 r^i_{klm} = -\gamma^i_{kl,m} + \gamma^i_{nl} \gamma^n_{km}
 + \gamma^i_{km,l} - \gamma^i_{nm} \gamma^n_{kl} \,.
 \ee
 Then the Ricci-like (but {\bf non-symmetric}) curvature tensor can
 be defined by contracting the indices $i, m$ (or, equivalently, $i, l$):
 \be
 \label{2}
 r_{kl} = -\gamma^m_{kl,m} + \gamma^m_{nl} \gamma^n_{km}
 + \gamma^m_{km,l} - \gamma^m_{nm} \gamma^n_{kl}
 \ee
 (let us stress once more that $\gamma^m_{nl} = \gamma^m_{ln}$
 but $r_{kl} \neq r_{lk}$). Using only these tensors and the
 anti-symmetric tensor density one can build up a rather
 rich geometric structure. In particular, Eddington discussed
 different sorts of tensor densities \cite{Ed}.
 A notable scalar density is\footnote{
 Following notation of Eddington we denote tensor densities by
 boldface Latin letters. Keeping clear distinction between tensor
 densities and tensors is important, especially,
 as far as there is no metric tensor  at our disposal.}
 \be
 \label{3}
 {\Lag} \equiv \sqrt{ -\det(r_{ij})} \,
 \equiv \, \sqrt{ -r} \,,
  \ee
 which resembles the fundamental scalar density of the Riemannian
 geometry, $\sqrt{-\textrm{det}(g_{ij})} \equiv \sqrt{-g}$.

   For this and some other
 reasons Eddington suggested to identify the
 symmetric part of $r_{ij}$,
 \be
 s_{ij} \equiv \half (r_{ij} + r_{j\,i}) \,,
 \label{3a}
 \ee
 with the metric  tensor $g_{ij}$.
 The anti-symmetric part,
 \be
 \label{4}
 a_{ij} \equiv \half (r_{ij} - r_{j\,i}) =
 \half (\gamma^m_{im,j}  - \gamma^m_{jm,i}) \,,
  \qquad  a_{ij,\,k} + a_{jk,i} + a_{ki,j} \equiv 0\,,
 \ee
 strongly resembles the electro-magnetic field tensor and it seemed
 natural to identify it with this tensor.
 Eddington tried to write consistent equations of
 the generalized theory
 but this problem was solved only by Einstein,
 with a different definition of the metric.

 The starting point for Einstein (in his first paper
   of the series  \cite{Einstein1})
 was to write the action principle and to suppose (\ref{3}) to be
 the Lagrangian density depending on 40 connection functions
 $\gamma^m_{kl}$.
 Varying the action w.r.t. these functions
 he derived 40 equations that allowed him to find the
 expression for $\gamma^m_{kl}$ that is given by (\ref{a3}) with
 $\alpha = - \beta = \third$.
 In the second paper he proved that this result is valid for
 any Lagrangian density depending only on $s_{ij}$
 and $a_{ij}$:
 \be
 {\Lag} \, = \, {\Lag} (s_{ij} , a_{ij}).
 \label{5}
 \ee
 The most elementary example is
 ${\Lag} (s_{ij} + \alpha a_{ij})$, where
 $\alpha$ is a dimensionless constant.

 Let us reproduce the main steps of the proof. Following
 Einstein we define the new tensor densities,
 $\textbf{g}^{ij}$,  $\textbf{f}^{ij}$, by the
 relations\footnote{
 Einstein's notation is somewhat casual because it tacitly
 assumes that $\textbf{g}^{ij}$ is symmetric and $\textbf{f}^{ij}$ -
 anti-symmetric while explicit derivations show that this is
 not automatically true. In fact, the derivatives
 in (\ref{10}) - (\ref{10a}) should be properly symmetrized and
 in concrete derivations it is not difficult.
  More important is to note that these definitions tacitly
 assume that geometry has just one dimensional constant,
 e.g., the cosmological constant $\Lambda$ having dimension
 $L^{-2}$. Possibly, the characteristic dimensions for the symmetric
 and antisymmetric parts of geometry are different and then,
 to restore the correct dimension in (\ref{10}), (\ref{10a}),
 we should multiply the densities $\textbf{g}^{ij}$,  $\textbf{f}^{ij}$
 by $\Lambda_s$, $\Lambda_a$ or, alternatively, introduce into
 geometry a new fundamental dimensionless constant, see next Section. }:
  \be
 \label{10}
  {{\p {\Lag}} \over {\p s_{ij}}} \equiv \textbf{g}^{ij} \,, \qquad
 {{\p {\Lag}} \over {\p a_{ij}}} \equiv \textbf{f}^{ij} \, ,
  \ee
  and also introduce a conjugate Lagrangian density,
  ${\Lag}^* \, = \, {\Lag}^* (\textbf{g}^{ij} , \textbf{f}^{ij})$
  by a Legendre  transformation
 \be
 \label{10a}
 s_{ij} = {{\p {\Lag}^*} \over {\p \textbf{g}^{ij}}} \,, \qquad
 a_{ij} = {{\p {\Lag}^*} \over {\p \textbf{f}^{ij}}} \,.
   \ee
   By varying ${\Lag}$ in $\gamma^i_{kl}$ and using
   the above definitions one can show that vanishing
   $\delta {\Lag} / \delta \gamma^i_{kl} = 0$ gives the
   following 40 equations
   \be
   \label{10b}
  2 \nabla^{\gamma}_i \, \textbf{g}^{kl} \,= \,\,
  \delta^l_i \, \nabla^{\gamma}_m \, (\textbf{g}^{km} + \textbf{f}^{km}) +
 \delta^k_i \, \nabla^{\gamma}_m \, (\textbf{g}^{lm} +
 \textbf{f}^{lm}) \,,
   \ee
 where $\nabla^{\gamma}_i$ is the covariant derivative w.r.t. to the
 affine connection $\gamma$. Using the expression for the covariant
 derivative of $\textbf{f}^{kl}$,
 \be
 \label{10c}
 \nabla^{\gamma}_i \, \textbf{f}^{kl} \,= \,\,
 \p_i \, \textbf{f}^{kl} + \,\gamma^k_{im} \, \textbf{f}^{ml} +\,
 \gamma^l_{im} \, \textbf{f}^{km}   -
 \gamma^m_{im} \, \textbf{f}^{kl} \, ,
 \ee
  we find an important relation allowing us to define the
  fundamental vector density $\textbf{a}^k$:
 \be
 \label{10d}
 \nabla^{\gamma}_i \, \textbf{f}^{ki} \,= \,\,
 \p_i \textbf{f}^{ki} \, \equiv \, \textbf{a}^k .
  \ee
 Then, from this relation and (\ref{10b}) we find that
 \be
   \label{10e}
  \nabla^{\gamma}_i \, \textbf{g}^{ik} \,= \,\,
  - {\frac{5}{3}} \textbf{a}^k  \,,
   \ee
 and thus
 \be
   \label{10f}
  \nabla^{\gamma}_i \, \textbf{g}^{kl} \,= \,\,
  - \third ( \delta_i^k \textbf{a}^l + \delta_i^l \textbf{a}^k).
   \ee
 Now we can complete Einstein's derivation of (\ref{a3}).
 To do this one has first to define the Riemann metric tensor
 $g_{ij}$. It is natural to define it by the equations
 \be
 \label{11}
 g^{kl} \sqrt{-g} \, = \, \textbf{g}^{kl} \,, \quad
 g_{kl} \, g^{lm} \, = \, \delta^m_k \,.
\ee
 Using $g_{ij}$ one can also define the covariant derivative
 $\nabla_i$ w.r.t. the Riemann connection so that
 \be
 \label{11a}
 \nabla_i \, g_{kl} = \nabla_i \, g^{kl} = 0 .
  \ee

 With these prerequisites, we can now use (\ref{10f}) to derive
 the expression  for $\gamma_{ij}^k$  in terms of the metric
 tensor $g_{ij}$ and of  the vector $a^k \equiv \ba^k/\sqrt{-g}$
 (and to find that in Eq.(\ref{a3}) $\alpha = - \beta = \third$):
 \be
 \gamma_{ij}^k  \,=\, \Gamma_{ij}^k + {\frac{1}{6}}(\delta_i^k a_j +
 \delta_j^k a_i) -\half g_{ij} a^k \,,
 \label{a31}
 \ee
 where $\Gamma_{ij}^k$ is the Riemann connection corresponding to
 the metric $g_{ij}$ (the Christoffel symbol).
 We omit this simple but
 tedious derivation which is very similar to finding the standard
 Riemann connection in general relativity (for more details
 see \cite{Einstein1}-\cite{Einstein2}).

 Returning to the definition of the nonsymmetric Ricci tensor
 $r_{ij}$ (\ref{2}) we can now derive $s_{ij}$, $a_{ij}$ in terms of
 $\gamma^k_{ij}$ (and thus in terms of the metric $g_{ij}$
 and of the vector field $a_k$):
 \be
 s_{ij} = R_{ij} + {\frac{1}{6}} a_i a_j \,, \qquad
 a_{ij} = {\frac{1}{6}} (a_{i,j} - a_{j,i}) ,
 \label{11b}
 \ee
 where $R_{ij} \equiv R_{ij}(g)$ is the standard Ricci tensor.

 Up to now we did not use any information on $\Lag$ except
 its dependence on $s_{ij}$ and $a_{ij}$ (correspondingly,
 ${\Lag}^*$ depends only on $\textbf{g}^{ij}$ and
 $\textbf{f}^{ij}$). To write the complete system of equations
 we should choose an analytic expression for the Lagrangian density.
 The  best candidate is of course Eddington's invariant (\ref{3}).
 Using the fact that this invariant is the second-order homogeneous \
 function of $r_{ij}$ we find that ${\Lag}^*$ is equal to
 ${\Lag}$. This means that it is sufficient to derive
 $\Lag$ as a function of $\textbf{g}^{ij} + \textbf{f}^{ij}$.
 It is clear (from (\ref{10})) that
 \be
 \textbf{g}^{ij} + \textbf{f}^{ij} \,=\,
 {\frac {-1}{2\sqrt{-r}}} \, \bar{r}^{ij}, \qquad
 \sum_k r_{ik} \, \bar{r}^{jk} \,\equiv \, r \,\delta_i^j .
  \label{11c}
 \ee
 Then it follows that $\det (\bar{r}^{ij}) = r^3$ and thus
 \be
 \det(\textbf{g}^{ij} + \textbf{f}^{ij}) \,=\,
 (2\sqrt{-r})^{-4} \, r^3 \,=\, r/16 \,.
   \label{11d}
 \ee
 Using this result one can find the final expression for
  $\Lag$ (and thus for ${\Lag}^*$)
 \be
 \label{11e}
 {\Lag} \equiv \sqrt{ -\det(r_{ij})} \,
 \equiv \, \sqrt{ -r} \,=\,
 4\sqrt{-\det(\textbf{g}^{ij} + \textbf{f}^{ij})}
 \, \equiv \, 4\sqrt{-\det(g_{ij} + f_{ij})} \,.
  \ee
 This Lagrangian was first proposed by Eddington who
 tried  (not quite successfully) to
 identify $s_{ij}$ with the metric tensor and $a_{ij}$ with the
 electromagnetic field tensor. Einstein gave a more correct
 interpretation of the intuitive Eddington idea and computed
 the same formula which has a different meaning.

 Using this result it is possible to complete the equations
 (\ref{11b}) by expressing the $s_{ij}$ and $a_{ij}$ in terms
 of $\textbf{g}^{ij}$ and $\textbf{f}^{ij}$ and finally to write
 equations for $g_{ij}$ and $f_{ij}$. In this report we will not
 go into this a bit cumbersome derivation and simply reproduce
 Einstein's {\bf effective Lagrangian} for the theory produced
 by the Lagrangian (\ref{3}):
 \be
 \label{8}
 {\Lag}_{eff} = -2 \Lambda \sqrt{-\det(g_{ij} + f_{ij})} \, + \,
 \sqrt{-\det(g_{ij})} \, \biggl[ R(g) + \sixth g^{ij} a_i a_j \biggr] \,,
 \ee
  which should be varied w.r.t. the metric and the
  vector field.\footnote{Here we modernize the original
  Einstein's notation to make a different interpretation of his
  theory obvious. The cosmological constant $\Lambda$ is introduced
  for keeping the correct dimensions: $[\Lambda] = [R]
  = \textrm{L}^{-2}$
  and $[a_i] = \textrm{L}^{-1}$. Dimensions which were completely neglected
  above will be restored in the next section.}

  The first term in the Lagrangian (\ref{8})
  is essentially nonlinear and singular
  This probably was one of the reasons to forget it for long time.
  Its version with Minkowski metric reappeared in a simpler context of
  `nonlinear electrodynamics' in 1934 \cite{Born}
  and nowadays is usually called the Born - Infeld
  Lagrangian.\footnote{
  In view of the above considerations (for a more detailed
  historical review see \cite{ATF}),
  calling the Lagrangians like (\ref{8}) the `Eddington -
  Born - Infeld' or even `Born - Infeld - Einstein'
  Lagrangians (see \cite{Deser}, \cite{Banados})
  is hardly justified.
  The models of these papers as well as
  some modern superstring ideas on inflation (see, e.g.,
  \cite{Langlois}) are much closer in spirit to the
  Eddington and Einstein ideas than to the beautiful
  BI nonlinear electrodynamics,
  that is close in spirit to ideas of G.~Mie \cite{Mie}.
  Note also that the main emphasis of \cite{Born} is on quantum
  theory and it is not discussing GR.}
  To the best of my knowledge the beautiful Einstein theory
  (\ref{8}) was not seriously studied (I could not even find
  a reference to it in papers considering related topics).
  From today's point of view it is of significant interest
  because it predicts dark energy
  (nonvanishing cosmological constant) and
  a new field - the massive  or tachyonic vector field coupled to
  gravity only.

 \section{Interpretation and generalization of the affine theory}
 The meaning of $g_{ij}$ and $R(g)$ in (\ref{8}) must be clear.
 To clarify the interpretation of $a_i$ and $f_{ij}$ let us expand
 the first term in powers of $f$
 up to the $f^2$-order terms. Generalizing (\ref{8}) by introducing
 a new dimensionless parameter $\lambda$, we easily find\footnote{
 Actually, we use this new parameter to
 disentangle the scale of the mass parameter of the vector field
 from the cosmological
 constant. We will see that for $\lambda = 1$, i.e. for the original
 Eddington - Einstein Lagrangian (\ref{11e}), the mass parameter will be
 close to $\sqrt{\Lambda}$.
 }
 \be
 \sqrt{|\det (g_{ij} + \lambda f_{ij})|} \,=\, \sqrt{-g} \,
 \sqrt{\det (\delta^i_j \,+ \lambda f^i_{\,j} )} \,=\,
 \sqrt{-g} \, \biggl(1 + \quart \lambda^2 f_{ij} f^{ij} + ...\biggr) ,
 \label{12}
 \ee
 and thus the approximate effective Lagrangian contains
 the cosmological constant term as well as the Proca
 Lagrangian \cite{Proca} of the real massive (tachyonic)
 vector field $a_i$ written several years
 after the Weyl and Einstein papers containing this theory as a part
 of the generalized gravity.

 However, we should
 emphasize that this is only the first approximation. The exact
 theory is much more complex. Indeed, one can see that the exact
 determinant may have zeroes for large enough `electric' vector
 field $\vec{e}_i \equiv f_{0i}$.
 For example, if $f_{ij} = 0$ for $i,j \neq 0$ we have\footnote{
 This configuration is realized in the spherically  symmetric case.
 Then the nonvanishing potentials are $a_t \equiv a_0(t,r)$,
 $a_r \equiv a_1(t,r)$ and the only nonvanishing field component is
 $f_{01}$
 }
  \be
 \det (\delta^i_j \,+ \lambda f^i_{\,j} ) =
  1 - \lambda^2 |\vec{e}^2| ,
 \label{12a}
 \ee
 where $\vec{e}^2 \equiv |g^{00}| \sum_i g^{ii} e_i^2$.
 Therefore, in the exact theory there is a nontrivial interplay
 between dark energy and dark matter. The existence of the upper
 bound on $|E|$ may give us a chance to establish the maximal
 ratio of dark matter to dark energy (supposing, of
 course, that the vector field gives a dominant contribution to
 the dark matter density). This will be possible only in
 a definite cosmological scenario presumably including inflation
 and some other sorts of matter. This motivates
  us at a certain  generalization of the model.

 Before we turn to this task, let us finish the interpretation of the
 basic Einstein models, which requires introducing physical fields with
 correct dimensions. To do this we just look into the third paper of
 Ref.\cite{Einstein1}, where Einstein writes the final equations for
 his second model\footnote{Einstein proposed this Lagrangian without
 relation to the nonlinear Lagrangian (\ref{11e}). Later he
 decided to take $\Lambda = 0$ with finite $\lambda^2 \Lambda$, but his final
 verdict was that this theory is also not describing `real physics'.
 },
 \be
 {\Lag}^* = -2\Lambda \sqrt{-g}
 \biggl(1 + \quart \lambda^2 f_{ij} f^{ij}\biggr) .
 \label{12aa}
 \ee
 We will see that $\lambda$ allows one to make the mass of the vecton
 absolutely arbitrary.

 Now, the equations of motion can be written by calculating
 $s_{ij}$ and $a_{ij}$ using  ({\ref{11b}) and (\ref{10a}):
 \be
 R_{ij} - \Lambda \, g_{ij} \,= -\lambda^2 \Lambda \,
 \biggl[f_{ik} f^k_{\,j} - \quart g_{ij} f_{kl} f^{kl} \biggl]
 - \sixth a_i a_j \,,
 \label{12b}
  \ee
  \be
  \Lambda \, f_{ij} =\, \sixth (\p_i a_j - \p_j a_i) ,
   \label{12bb}
  \ee
  and the equation for the potential $a_i$ immediately follows from
  (\ref{10d}) and (\ref{12bb}). Here $g_{ij}$, $f_{ij}$ are dimensionless
  while $[a_i] = \textrm{L}^{-1}$ and we use the system in which
  $c = 1$. Now, let us multiply $a_i^2$ by some $c_A^2$ having the dimension
 $\textrm{M L}^3 \textrm{T}^{-2}$ and define the standard fields
 $F_{ij}$, $A_i$ having the correct CGS  dimensions:
 \be
 6 \, c_A \, \Lambda \, f_{ij} \equiv F_{ij} =\, \p_i A_j - \p_j A_i \,,
 \qquad  A_i \equiv c_A a_i \,.
 \label{12c}
 \ee
   Rewriting Eq.(\ref{12b}) in terms  of the physical fields we find:
 \be
 R_{ij} - \Lambda \, g_{ij} \,= -\kappa  \, \biggl[F_{ik} F^k_{\,j} -
 \quart g_{ij} F_{kl} F^{kl} - \mu^2 A_i A_j \biggr] \,,
 \label{12d}
 \ee
 where $\kappa^{-1} \equiv 36 \, \lambda^{-2} \Lambda \, c_A^2$ and
 $\mu^2 \equiv -6 \, \lambda^{-2} \Lambda$. Adjusting the parameter
 $c_A$ we can identify $\kappa$ with the gravitational constant $G$
 (in what follows we also take $\kappa = 1$)
 and, identifying $\kappa \equiv G/c^4$, we have standard notation.
 These equations as well as the equations for
 $A_i$ can be derived from the {\bf effective Lagrangian},\footnote{
 Note that, according to Eq.(\ref{12b}), $\mu^2$ is negative and
 thus the vecton must be a tachyon. In two first papers of
 \cite{Einstein1}, Einstein apparently assumed that $\mu^2 > 0$ and
 in our paper \cite{ATF} we called the Einstein model the
 corresponding effective Lagrangian with the positive $\mu^2$.
 Thus the {\bf geometric} Einstein model does not completely
 correspond to his {\bf effective dynamical model}. }
 \be
 {\Lag}_{eff} \, = \,  \sqrt{-g}\,
 \biggr[R - 2\Lambda - \kappa \biggr(\half F_{kl} F^{kl}+
 \mu^2 A_k A^k \biggl) \biggl]\,,
 \label{13}
 \ee
 by varying it in $g_{ij}$ and $A_i\,$. Quite similarly one can write
 the effective Lagrangian for the exact `square-root' Lagrangian.
 In fact, one can simply restore the square root from the second and
 third terms  in the approximate Lagrangian (\ref{13}) and we leave
 this exercise  to the reader.\footnote{
 The  approximate Lagrangian  (\ref{13}) correctly describes   properties
 of any affine theory with one vector field  in the limit of small deviation
of the affine connection  from the  Riemannian one. In the next
order  in $A^2$ (not speaking  of the strong - field limit)  the
relevant physical  model will essentially  depend on the concrete
structure  of the chosen affine geometry. }

 Here it is appropriate to say a few words about the free parameters
 in the Lagrangian
 (we only discuss the case of the positive $\mu^2$). The present
 value of $\Lambda$ is estimated as $\simeq 10^{-56} \, \textrm{cm}^{-2}~$
 (see, e.g., \cite{Starobin} - \cite{Rubakov}).
 The de Broglie length of the vecton in smallest galaxies must be less
 than their dimension $\sim 10^{21}\textrm{cm}$. If the average velocity of the
 vecton is $\sim 10^{-3}c$ we find that
 $\lambda \, {\Lambda}^{-\half} < 10^{-3} \, 10^{21} \sim 10^{18} \textrm{cm}$
 and thus we must
 suppose that $\lambda < 10^{-10}$. Such a naive estimate may be criticized
 and we do not insist on it.\footnote{
  If we do not wish to introduce additional very large/small
   dimensionless parameters and take $\lambda \sim 1$, the estimated mass of
 the vecton will be much less than the best present lower bound on the photon mass
 ($\leq 10^{-51} \textrm{g}$, see \cite{gamma}) and, strictly speaking,
 the  interpretation of the vecton as a massive photon cannot be
 completely excluded without further, more serious considerations.}
 For these reasons, in what follows we consider the parameters as free.
 In cosmological applications, this is most natural in the context of the
 {\bf multiverse approach} (see, e.g., discussions in \cite{multiverse}).
 The present status of dark energy, dark matter and
 inflation does not allow us to make more precise statements about the values
 of the parameters and we proceed with purely mathematical formulation not
 returning to quantitative estimates.

 The most popular theories of  {\bf inflation}, require a few
 {\bf massive scalar mesons} (see, e.g., \cite{Starobin} - \cite{Rubakov}).
 In the frame of the ideas discussed above there may emerge several
 inflation mechanisms. For the approximate model (\ref{13}) with
 $\mu^2 > 0$ there is no obvious inflation mechanism.
 In our study of a simplified cosmological
 model based on the vecton gravity (see \cite{ATF}) it was shown that in
 spite of the fact that there exist some simple exponentially expanding
 solutions they hardly can be used in any inflationary scenario.
 On the other hand,  a reasonable  inflationary type solution
 is possible  for $\mu^2 < 0$ (see \cite{Ford}).
 Perhaps, it is easier to find a realistic inflation
 in gravity coupled to many massive vector mesons (see
 \cite{Bertolami} - \cite{Germani}) but most natural and transparent
 models of inflation use massive scalar mesons (usually
 called inflatons). It is of interest to note that the most
 natural geometric theory with the square - root Lagrangian
 strongly resembles some recently studied `Born - Infeld' models
 for inflation obtained in the frame of string theory
 (see, e.g. \cite{Langlois}). Although the geometric Einstein
 Lagrangian (even $\lambda$-modified) predicts negative $\mu^2$,
 it is possible that for a more general connection (\ref{a3})
 this parameter is positive. We briefly discuss this possibility
 in Appendix.

 We now demonstrate how such massive (or, tachyonic) scalar particles
 may be produced by the simplest
 dimensional reduction of a {\bf higher dimensional generalization} of the
 Einstein affine model. In fact, the previous construction can
 easily be extended from the dimension $D=4$ to any dimension $D \neq 2$.
 It easy to find that we must make the following replacements in the
 geometric equations:
 in Eq.(\ref{10e}) we replace $5/3$ by $(D+1)/(D-1)$, in Eq.(\ref{10f})
 -- $1/3$ by $(D-1)^{-1}$, and Eqs.(\ref{a31}), (\ref{11b})
 are replaced by:
 \be
  \gamma^m_{kl} = \Gamma^k_{ij} +
 [(D-1)(D-2)]^{-1} [\, \delta^m_k \, a_l + \delta^m_l \, a_k -
   (D-1) g_{kl}\, a^m ] \biggr] \,,
 \label{a31a}
 \ee
 \be
 s_{ij} = R_{ij}(g) + [(D-1)(D-2)]^{-1} \, a_i a_j \,, \qquad
  a_{ij} = [(D-1)(D-2)]^{-1} \, (a_{i,j} - a_{j,i})\,.
 \label{11ba}
 \ee
 The equations related to dynamics, starting from (\ref{11d}),
 depend on $D$ more significantly, as can be seen from the
 relation that can easily be derived:
 \be
 {\Lag}_{eff} \, \equiv \, \sqrt{ -\det(r_{ij})} \, \sim
 \, \sqrt{-g} \ [\det (\delta_i^j + f_i^j)]^{1/D-2} \,.
 \label{dr}
 \ee
 (Note that this expression is meaningless when $D=2$.)
 Now, by making the {\bf simplest dimensional reduction}
  to $D=4$,
  we can interpret the components of the vector field $a_k$ with
  $k \geq 4$  as real massive scalar fields\footnote{
  From now on we use the term `massive' both for real and
  imaginary (tachyonic) masses. In Appendix we hint at the
  possibility of geometric models with real masses of the vecton
  and scalar particles.}
 the geometric masses of which coincide with the vecton mass.
 The components $F_{ik}$ ($k \geq 4, \, i \leq 3$) give the
 kinetic terms of the scalar fields $A_k (x^0, x^1. x^2, x^3)$.
 In the exact Lagrangian (\ref{8}), $\det(g_{ij} + f_{ij})$
 will depend both on derivatives of the vecton and and of
 the scalar fields. In the $F^2$-approximation (\ref{13}),
 we get the standard scalar terms. It is sufficient to
 write this expression with  one scalar field:
 \be
 {\Lag}_{eff}\, = \,  \sqrt{-g}\,
 \biggr[R - 2\Lambda -
 \kappa \biggr(\half F_{kl} F^{kl} + \mu^2 A_k A^k
 + g^{ij} \p_i \psi \, \p_j \psi + m^2 \psi^2 \biggl) \biggl]\,.
 \label{13a}
 \ee
 In next Section  we discuss only this linearized (in $A_k$
 and $\psi$) model, which we further simplify by considering
 its spherically symmetric sector.\footnote{
 The spherical reduction can equally be applied to the
 effective Lagrangian (\ref{8}) and, with some caution, to the original
 Lagrangian (\ref{3}).
  }

 \section{Spherical symmetry and cosmology}
 In general,  the spherically reduced theory
 is described by two-dimensional differential equations which
 are not integrable except very special cases
 (for many examples and references see, e.g.,
 \cite{CAF1}-\cite{ATF7}).
  Following the approach to dimensional reduction and to
 resulting 1+1 dimensional dilaton gravity (DG)
  developed in papers
  \cite{CAF1}-\cite{ATF4}, \cite{ATF1}, \cite{ATF2}-\cite{ATF7}
 it is not difficult to derive these equations and the reader
 can find them in \cite{ATF}.

 The dilaton gravity coupled to the massive vector field
 (I suggest to call it {\bf vecton-dilaton gravity, VDG})
 is more complex than the well
 studied models of dilaton gravity coupled only to scalar fields and
 thus it requires a separate study.
  The  first thing to do is to further reduce the theory to static
 or cosmological configurations.

  The simplest way to derive the corresponding equations is to suppose
 that all the functions in the equations depend either on  $r$ or on
  $t$. But this
  is not the most general dimensional reduction of
  the two-dimensional theory.
 There exist more general ones that allow one to simultaneously treat
 static states (black holes), cosmologies and some waves.
 These generalized reductions were proposed in papers
 \cite{ATF5}, \cite{ATF3}, which were
  devoted to dilaton gravity coupled to scalar fields and Abelian gauge
  fields. Here we only discuss the cosmological
  reductions (for more detailed presentation of static and
  cosmological solutions see \cite{ATF}).

   The simplest cosmology
 can be obtained by the same naive reductions as was used for
 static states in \cite{ATF}). However, this is not the most general dimensional
 reduction giving all possible spherically symmetric cosmological
 solutions.
 A more general procedure is described in \cite{ATF3}.
 Following this procedure we write the general spherically
 symmetric metric as:
 \be
 ds_4^2 = e^{2\alpha} dr^2 + e^{2\beta} d\Omega^2 (\theta , \phi) -
 e^{2\gamma} dt^2 + 2e^{2\delta} dr dt \, ,
 \label{eq1}
 \ee
 where  $\alpha, \beta, \gamma, \delta$ depend on $t$, $r$ and
 $d\Omega^2 (\theta , \phi)$ is the metric on the 2-dimensional sphere
 $S^{(2)}$. Then  the two-dimensional reduction of the four-dimensional
 theory (\ref{13a}) can easily be found
 (here the prime denotes differentiations in $r$ and the dot -
 in  $t$:
 \be
 \cL^{(2)} = e^{2\beta} \bigl[e^{-\alpha - \gamma} (\dA_1 - A_0')^2 -
 e^{-\alpha + \gamma} (\psi'^2 + \mu^2 A_1^2) +
 e^{\alpha - \gamma} (\dpsi^2 + \mu^2 A_0^2) -
 e^{\alpha +\gamma} (V + 2\Lambda)  \bigr] + \cL_{gr} \,,
  \label{eq2}
 \ee
 where $\psi = \psi(t,r)$, $V=V(\psi)$, $A_i = A_i(t,r)$,
 $\dA_1 - A_0' \equiv F_{10}$ and
 \be
 \cL_{gr} \equiv
 e^{-\alpha + 2\beta +\gamma} (2\beta'^2 + 4\beta' \gamma') -
 e^{\alpha + 2\beta - \gamma} (2\dbe^2 + 4\dbe \dal) +
  2k e^{\alpha + \gamma}
 \label{eq3}
 \ee
 is the gravitational Lagrangian, up to the omitted total derivatives
 that do not affect the equations of motion.
 Variations of this Lagrangian give all the equations of motion except
 one constraint,
 \be
 -{\dot{\beta}}^{\prime} - \dot{\beta} {\beta}^{\prime} +
  \dot{\alpha} {\beta}^{\prime} + \dot{\beta} {\gamma}^{\prime} \,\,
  = \,\, \half [\dot{\psi} {\psi}^{\prime} + A_0 A_1] ,
 \label{eq4}
 \ee
 which should be derived before we omit the $\delta$-term in the metric
 (taking the limit $\delta \rightarrow -\infty$). All other equations
 of motion can be obtained from the effective Lagrangian
 (\ref{eq2}).

 Now, the distinction between {\bf static} and {\bf cosmological}
 solutions is in the dependence of their `matter' fields $A_i$ and $\psi$
 on the space-time coordinates.
 We call {\bf static} the solution for which $A_i=A_i(r)$ and $\psi=\psi(r)$.
 If $A=A_i(t)$ and $\psi=\psi(t)$ we call the solution {\bf cosmological}.
 There may exist also the {\bf wave-like} solutions for which
 $A$ and $\psi$ depend on linear combinations of $t$ and $r$ but here
 we do not discuss this possibility (see, e.g., \cite{ATF7} and references
 therein). For both the static and cosmological solutions the gravitational
 variables in general depend on $t$ and $r$.
 To solve the equations of
 motion we may reduce them by separating $t$ and $r$.
 It is clear that to separate the variables $r$ and $t$ in the metric we
 should require that
 \be
 \alpha = \alpha_0(t) + \alpha_1(r) , \quad \beta = \beta_0(t) +
 \beta_1(r) , \quad \gamma = \gamma_0(t) + \gamma_1(r) , \,
 \label{eq5}
 \ee
 Inserting this into the equations of motion one can find the restrictions
 on the gravitational (and, possibly on the matter) variables that
 must be fulfilled. The details can be found in \cite{ATF3}, where one can
 find the complete list of the static and cosmological
 spherically symmetric solutions when
 the vector field identically vanishes (we call this case the `scalar cosmology).
 Here we only give a very brief summary
 and a simple generalization to nonvanishing vector field.

 The naive cosmological reduction (that supposes all the fields to be
 independent of $r$) does not give the standard FRW scalar cosmology.
 As was shown in \cite{ATF3}
 (see also the earlier paper \cite{ATF1}),  all homogeneous isotropic
 cosmologies should satisfy the following conditions:
 \be
 \label{eq6}
 \dot{\alpha} = \dot{\beta} \,, \quad \gamma' = 0 \,, \quad
 {\beta_1}'' + k e^{-2 \beta_1} = 0 \,, \quad
     k e^{-2 \beta_1} \,-\, 3{\beta_1'}^2 - 2{\beta_1}'' \,=\, C \,,
 \ee
 where $C$ is a constant proportional to the 3-curvature
 (its time dependence is given by the factor $e^{-2 \alpha_0}$)
 and the third equation is the isotropy condition.
 Neglecting inessential constant factors, we also have chosen
 $\alpha_1 = \gamma_1 = 0$. We see that for the naive reduction
 the isotropy conditions in (\ref{eq6})
 can be satisfied only if $k = 0$ and
 that the first condition is not dictated by (\ref{eq4}).
 Thus, the naive reduction gives, in general, a homogeneous
 non-isotropic cosmology.

 For the FRW {\bf scalar cosmology} $\beta' \neq 0$ and all the
 conditions (\ref{eq6}) must be satisfied. Then the effective
 one-dimensional Lagrangian describing both the naive and FRW cosmology is
 \be
 \cL^{(1)} = 6\bar{k} e^{\alpha + \gamma} -
 e^{2\beta} \bigl[e^{\alpha +\gamma} (V + 2\Lambda)  -
 e^{\alpha - \gamma} (2\dbe^2 + 4\dbe \dal - \dpsi^2) \bigr] \,,
 \label{eq7}
 \ee
 where $\bar{k}$ is a new real constant related to $k$ and $C$;
 $\alpha, \beta, \gamma$ and
 $\psi$ depend only on $t$.
 Taking $\alpha(t) = \beta(t)$ we get
 the Lagrangian of the FRW scalar cosmology, for which
 it is not difficult to derive the equations of motion.

 We now write the general cosmological `Wedein' Lagrangian supplemented
 by the minimally coupled scalar field (that may represent either
 matter or inflaton). At first sight, the dimensional reduction of
 the spherically symmetric Lagrangian (\ref{eq2})-(\ref{eq3})
 with the vector field must not differ from the usual one used for
 the scalar cosmology and can be written as:
 \be
 \cL^{(1)} = 6\bar{k} e^{\alpha + \gamma} +
 e^{2\beta} \bigl[e^{-\alpha - \gamma} \dA_1^2 -
 e^{-\alpha + \gamma} \mu^2 A_1^2 -
 e^{\alpha +\gamma} (V + 2\Lambda)  -
 e^{\alpha - \gamma} (2\dbe^2 + 4\dbe \dal - \dpsi^2) \bigr] \,,
 \label{eq10}
 \ee
 where all the fields depend on $t$.
 Then, taking $\alpha = \beta$, we apparently obtain
 a FRW type cosmology with the vector field.
 However, unlike the scalar field,
 the two-dimensional vecton field equations,
 \be
 \label{ac4}
 \d_{\,0} [e^{\alpha_0 -\gamma_0 + 2\beta_1} \dA_1 ]
 = -\mu^2 e^{\alpha_0 + \gamma_0 + 2\beta_1} A_1 \,,
 \ee
 \be
 \label{ac5}
 \ \d_{\,1} [e^{\alpha_0 -\gamma_0 + 2\beta_1} \dA_1 ]
 = -\mu^2 e^{3\alpha_0 - \gamma_0 + 2\beta_1} A_0 \,.
 \ee
  {\bf do give additional constraints} on $\beta_1(r)$.
 The first equation does not depend on $\beta_1$, but
 the second one requires either $\beta_1'(r) = 0$
 or $\beta_1'(r) = \textrm{const}$. The second condition
 gives $A_0 \sim \dA_1$ and so (\ref{eq4}) is incompatible
 with the isotropy condition $\dot{\alpha} = \dot{\beta}$.
  Therefore, there remains only the first case,
 $\beta_1'(r) \equiv 0$, from which it follows that
 $k  = \bk = 0$. Although the constraint (\ref{eq4}) is
 identically satisfied (as we suppose that $\gamma' = 0$)
 it does not give the necessary isotropy condition
 $\dot{\alpha} = \dot{\beta}$
 that automatically emerges in the scalar cosmology case.
 As we'll see in a moment, this condition cannot be
 exactly satisfied in the vecton cosmology and can only be
 approximate.

 Summarizing this discussion, we consider {\bf the vecton plus
 scalar cosmology} described by the Lagrangian (\ref{eq10})
 with $k=\bk =0$ and $A_1$ being the $A_z$ component of the
 four-dimensional vector field (it follows that the
 cosmology must be in general non-isotropic).
 To write the corresponding equations of motion in a
 most clear and compact form we introduce the temporal
 notation
 \be
 \rho \equiv \third (\alpha + 2\beta) \,, \quad
 \sigma \equiv \third (\beta - \alpha) \,, \quad
 A_{\pm} = e^{-2\rho + 4\sigma} (\dA^2 \pm
 \mu^2 e^{2\gamma} A^2) \,, \quad
 \bV \equiv V(\psi) + 2\Lambda \,.
 \label{eq11}
 \ee
 where $A_1 \equiv A_z \equiv A$. Then the exact
 Lagrangian for vecton plus scalar cosmology is:
 \be
 \cL^{(1)} = e^{2\rho - \gamma} (\dpsi^2 - 6\drho^2 + 6\dsig^2) +
 e^{3\rho - \gamma} A_{-}  -
 e^{3\rho + \gamma} \, \bV(\psi) \,.
 \label{eq12}
 \ee
 We see that $A, \psi, \rho, \sigma$ are the dynamical variables
 and $e^\gamma$ is the Lagrangian multiplier, variations of which
 give us the remaining energy constraint:
 \be
 \dpsi^2 - 6\drho^2 + 6\dsig^2 + A_{-} + e^{2 \gamma} \, \bV = 0
 \label{eq13}
 \ee
 (the momentum constraint
 (\ref{eq4}) is satisfied by construction). The other equations are:
 \be
 \ddot{A} + (\drho + 4\dsig - \dga) \dA +
 e^{2 \gamma} \mu^2 A = 0 \,,
 \label{eq14}
 \ee
  \be
 4\ddot{\rho} + 6\drho^2 -4\drho \dga  - 6\dsig^2 + \third A_{-}
 + \dpsi^2 - e^{2 \gamma} \, \bV = o\,,
 \label{eq15}
 \ee
 \be
 \ddot{\sigma} + 3 \dsig \drho - \dsig \dga -\third A_{-} = 0 \,.
 \label{eq16}
 \ee
 \be
 \ddot{\psi} + (3\drho - \dga)\dpsi +
 \half e^{2 \gamma} \, \bV_{\psi}  =  0 \,,
 \label{eq17}
 \ee
 These equations are much more complex than the equations
 of the scalar cosmology. They are {\bf not integrable}
 in any sense and rather difficult for a qualitative analysis.

 They would be greatly simplified if it were possible to
 neglect the $\sigma$-field. Unfortunately, it is evident that
 this is in general impossible because then $A_{-} =0$ and the last
 condition is incompatible with the other equations.
 This means that the exact solutions of the `Wedein' model (even with
 many scalar fields minimally coupled to gravity)
 should be {\bf non-isotropic}\footnote{
 If one would introduce other scalar fields non-minimally coupled to
 gravity, this statement may become not valid. At this stage of
 investigation, we are not ready to add other vector fields or
 fields with the the spin $1/2$.}.

 To get a {\bf simplified model} (`by brute force') we may neglect Eq.(\ref{eq17})
 and take $\sigma \equiv 0$, $\psi \equiv 0$, $V(\psi) \equiv 0$.
 Then  $\rho = \alpha = \beta$ and the approximate effective Lagrangian
 (\ref{eq12}) becomes\footnote{
 Above, we usually neglected the dimensions of all the variables
 and omitted the gravitational constant.
 Here we only need to restore one of the dimensions supposing that
 $[t^{-2}] = [k] = [\Lambda] = [\mu^2] = [L^{-2}]$. }
 \be
 \label{26}
 \Lag_a = - 6 \dal^2 e^{3\alpha - \gamma} -
 2\Lambda e^{3\alpha + \gamma} + \dA^2 e^{\alpha - \gamma}
 - \mu^2 A^2 e^{\alpha+\gamma} \,,
 \ee
 The corresponding equations of motion are the three equations
 (\ref{eq13})-(\ref{eq15}) with $\sigma = \psi = V =0$ and
 $\rho = \alpha$. The first equation, (\ref{eq13}), is
 equivalent to vanishing of the Hamiltonian.
 Denoting $f \equiv e^{\alpha}$ and taking
 the gauge fixing condition $\gamma = 0$,
 the {\bf `standard' gauge},  we have
 \be
 \label{27}
 \Hag^a_0 \equiv f \bigl[-6 \df^2 +2\Lambda f^2 +
  \dA^2 + \mu^2 A^2 \bigr] = 0 \,.
 \ee
 Another useful gauge, the {\bf LC gauge}, is
 $\alpha = \gamma$.
 In this gauge, the effective Hamiltonian is:
 \be
 \label{28}
 \Hag^a_1 \equiv -6 \df^2 + 2\Lambda f^4 +
 \dA^2 + f^2 \mu^2 A^2 = 0 \,.
 \ee
 It is also not difficult to write the equations independent
 of the gauge choice and we leave this as a simple exercise
 to the reader.

 In \cite{ATF} we constructed solutions of these equation
 using and iteration schemes. Recently, S.~Vernov derived some
 analytically exact solutions for special values of the parameters.

 \section{Additional remarks and discussion}
 To confront the affine theory to `real physics'
 we should first formulate and study in some detail
 the `really geometric' theory with the square-root Lagrangian.
 I was not completely satisfied with the introduction
 of the $\lambda$ parameter into this Lagrangian to solve the vecton
 mass problem, but recently I realized that there exists a much more
 beautiful solution. As was noted in the Appendix to the book
 \cite{Ed},
 for the non-symmetric matrix $r_{ik}$ {\bf there exists another
 scalar density similar to (\ref{3})}. If we replace $\det{r_{ik}}$
 by the following scalar density of the weight two
 (we may call it a `twisted' determinant),
 \bdm
 {\det}^{\prime}(r_{ij}) \equiv \frac{1}{4!} \epsilon^{ijkl}
 \epsilon^{mnrs}  r_{im} r_{jn} r^T_{kr} r^T_{ls} \,
 \equiv \frac{1}{4!} \,
 \epsilon \cdot r \cdot r \cdot r^T \cdot r^T \cdot \epsilon ;
 \qquad r^T_{kr} = r_{rk} \,,
 \edm
  we can write a more general Lagrangian:
 \be
 \label{new}
 {\Lag} \equiv \alpha \, \sqrt{ -\det(r_{ij})} +
 \alpha^{\prime} \, \sqrt{ -{\det}^{\prime}(r_{ij})} \,.
  \ee
 For this Lagrangian, the cosmological constant and the mass of
 the vecton can be made arbitrary.

 In fact, in the four-dimensional theory there exists one more
 scalar density of the weight two,
 \bdm
 {\det}^{\prime \prime}(r_{ij}) \equiv \frac{1}{4!} \, \epsilon^{ijkl}
 \epsilon^{mnrs}  r_{im} r_{jn} r_{kr} r^T_{ls}
 \equiv \frac{1}{4!}
 \epsilon \cdot r \cdot r \cdot r \cdot r^T \cdot \epsilon  \,,
 \edm
 the square root of
 which could also be added to the `geometric' Lagrangian (\ref{new}).
 All the three densities are equally `fundamental' from the geometric point
 of view but their dependence on the symmetric and antisymmetric
 parts of $r_{ij}$ is significantly different. Indeed, the density
 \bdm
 {\det}^{\prime \prime}(r_{ij}) \,=\, \det(s_{ij}) \,-\,
 \det(a_{ij}) \,
 \edm
 has no terms quadratic in $a_{ij}\,$ while $\det(r_{ij})$ and ${\det}^{\prime}(r_{ij})$
 contain the $a^2$-order terms of opposite sign.
  These terms are easy to derive for the diagonal matrix $s_{ij}$.
 Indeed, suppose that $s_{ij} = \delta_{ij} s_i$ and introduce
 for the matrix $a_{ij}$ the natural notation ($i = 1,2,3$):
 \bdm
 e_i = a_{0i}\,, \qquad {\tilde{e}}_i \,
 \equiv \, e_i / \sqrt{s_0 s_i} \,,
 \edm
 \bdm
 h_i \equiv \epsilon_{ijk} \, a_{jk} \,, \qquad
 {\tilde{h}}_i \, \equiv h_i / \sqrt{s_j s_k} \,.
\edm
 Then it is easy to derive the intuitively clear expressions
 for the three densities:
 \bdm
 \det (r_{ij}) =   [1 \,-\, {\tilde{e}}^2 \,+\,
 {\tilde{h}}^2 \,+\,
 (\tilde{e} \cdot \tilde{h})^2 ] \prod_i s_i \,\,,
 \qquad {\det}^{\prime \prime} (r_{ij}) =  [1 \,-\,
  (\tilde{e} \cdot \tilde{h})^2 ] \,\prod_i s_i \,\,,
 \edm
 \bdm
 {\det}^{\prime} (r_{ij}) =  [1 \,+\,
 \third {\tilde{e}}^2  \,-\, \third {\tilde{h}}^2 \,+\,
 (\tilde{e} \cdot \tilde{h})^2 ] \,\prod_i s_i \,\,.
  \edm

   A generalization of the Lagrangian (\ref{new}) containing
 three square  roots would define a rather complex
 and not very beautiful theory.
 An essentially equivalent but simpler Lagrangian can be made of
 the following  three scalar densities of the weight two
 ($a$ symbolizes the matrix $a_{ij}$):
 \be
 d_0 \equiv {\det} (s_{ij}) \,, \qquad
 d_2 \equiv  \epsilon \cdot s \cdot s \cdot a \cdot a \cdot \epsilon \,,
 \qquad
 d_4 \equiv \epsilon \cdot a \cdot a \cdot a \cdot a \cdot \epsilon \,.
 \label{den}
 \ee
 Then the geometric Lagrangian density,
 \be
 \label{new1}
 {\Lag} \equiv  \alpha \, \sqrt{ | \, d_0 | \,+
 \, \alpha^{\prime} d_2 \,+\,
 \alpha^{\prime \prime} d_4 } \,\,,
  \ee
 is conceptually as good as the Einstein's one, although
 it is more difficult to work with.
 Depending on the signs of the numerical coefficients one
 could then obtain positive or negative cosmological constant,
 as well as the standard or exotic (phantom) sign of the vector
 field kinetic energy. Note that, in general, the Lagrangian
 has zeroes, like the simpler Lagrangian of Born and Infeld.

 \bigskip
 Let us {\bf summarize} the main results, their possible generalizations,
 and applications. Einstein's approach to constructing the
 generalized theory of gravity consists of two stages.
   At the first  stage, it is
   supposed that the general symmetric connection should be
  restricted by applying the Hamilton principle to a general Lagrangian
 density depending either on $r_{ij}$ (the first two papers)
 or separately on $s_{ij}$ and $a_{ij}$ (the third paper).\footnote{
 This idea was quite alien to Weyl and Eddington who
 started from formulating a particular geometry.
 Thus they postulated the connection (\ref{a3}) with
 $\alpha = 1$, $\beta = 0$   and then tried to
 write some equations generalizing the Einstein equations.
 Although they differ with Einstein in the general approach and
 in the connection coefficients, the equations have many
 features in common, e.g., the nonvanishing cosmological
 constant (exactly or in some approximation) and massive/tachyonic
 vector field. }
 He gave no motivation for this Ansatz but it is easy to see
 that, in the limit $a_{ij}=0$,  the resulting theory is
 consistent with the standard general relativity with a cosmological
 term. At this stage Einstein succeeded in
 deriving the remarkable expression (\ref{a3}) for the connection
 and the general equations (\ref{11b}) that introduce into the
 play a (massive/tachyonic) vector field $a_i$ (at the same time,
  he naturally introduced the metric $\textbf{g}^{ij}$).

 At the next stage
 one should choose a concrete Lagrangian density
 ${\Lag}(s_{ij} \, , a_{ij})$.
 Einstein did not formulate any principle for selecting
 a Lagrangian, and from both geometrical and physical point
 of view his concrete choice looks sufficiently arbitrary,
 especially in the third paper.

 Let us try to formulate
 properties we consider natural for the geometric Lagrangian
 density  ${\Lag}$:
 {\bf 1.}~It must not depend on any dimensional constant.
 {\bf 2.}~Its integral over the $D$-dimensional space - time must be
 dimensionless. {\bf 3.}~Its analytic form is independent of $D$.
 {\bf 4.}~It should depend on tensors that have a direct geometrical
 meaning and a natural physical interpretation.
 {\bf 5.}~The most important requirement is that the
 resulting generalized theory must agree with the well
 established experimental consequences of the standard
 Einstein theory.

 The last property is rather difficult to check without
 a detailed study of the theory.
  The fourth condition
 is somewhat vague and depends of our
 understanding of what is `geometry' and what is `physics'.
 Clearly, the variables  $r_{ij}$,  $s_{ij}$,  and $a_{ij}$
 satisfy the condition 4, and the Lagrangian densities
 (\ref{3}), (\ref{new}), (\ref{new1}) satisfy the conditions
 1-4.  However, the vector field $a_i$ is also fundamental
 (see Appendix) so that we can construct a density depending also
 on it and satisfying the first four conditions. For example,
 let us define one more density of the weight 2:
 \bdm
 d_1 \equiv
 \epsilon \cdot s \cdot s \cdot s \cdot {\bar a} \cdot \epsilon \,,
 \edm
 where ${\bar a}$ symbolizes the matrix $a_i a_j\,$. Then, replacing
 in (\ref{new1}) the density $d_4$ by $d_1$, we obtain a Lagrangian
 satisfying the conditions 1-4 and containing an additional
 mass term (it is obvious if we take a diagonal matrix $s_{ij}$).

  One more question to Einstein's approach is about the role of
  the  metric tensor  $g_{ij}$ in geometry and in physics.
  More generally,  this is the question
  about the meaning of Einstein's  geometric Lagrangian.
 In  Weyl's  geometric approach,  a metric tensor  is introduced from the very
 beginning but is defined   up to a Weyl   transformation.
 In Einstein's approach, it is (seemingly uniquely) defined by
 the  Hamilton principle.  However, we  know that it depends
 on  our choice of the Weyl  frame  and  that
 the vector field also   depends on this choice.
 Einstein tacitly  bypassed this question  but we must  try to get a deeper
 understanding  of an interrelation between geometry  and  physics
 at this level. In particular,  the role of the conformal transformation,
 of the Weyl frame choice  and, especially, the significance
  of choosing different  independent fields in the
 geometric Lagrangian  must be carefully investigated  and understood.

 \section{Appendix}
 Here we give a very brief summary of the  main geometric
 facts used in his paper.
 As far as the author can judge, the best available introduction
 in the non-Riemannian geometry (for physicists) is the
 old book by Eisenhart \cite{Eisen}. Here we will speak about
 symmetric connections only, although many results are true
 or can be easily generalized to non-symmetric
 connections\footnote{
 The earliest explorers of non-symmetric connections
 in relation to gravity theory were Eddington, E.Cartan,
 and Einstein. A concise and beautiful introduction into
 this subject was
 given by E.Schr\"{o}dinger \cite{Erwin}. At present,
 one can see a revival of the interest to this field,
 see, e.g., \cite{Deser1} - \cite{Nair} and references therein.}.
 The general symmetric connection can be written as:
 \be
 \label{app1}
 \gamma_{ij}^k  \,=\, \Gamma_{ij}^k + A^i_{jk} \,,
 \ee
 where $\Gamma_{ij}^k$ is the Christoffel symbol for
 some $g_{ij}$ and  $A^i_{jk}$
 is an arbitrary tensor symmetric in the lower indices
 (more precisely, for any symmetric connection there exist
 a symmetric tensor $g_{ij}$ and a tensor
 $A^i_{jk} = A^i_{kj}$ such that (\ref{app1}) is valid).
 Defining the vector $A_k \equiv A^i_{ik}$ one can find that
 the antisymmetric part of $r_{jk}$ is equal to the rotor
 of $A_k$, i.e
 \be
 a_{ij} = \half (A_{i,j} - A_{j,i}) \,.
 \label{app2}
 \ee
 This equation, up to normalization, coincides with
 the second equation in (\ref{11b})
 that was derived above in a rather complex way.
   It follows that the
 fundamental vector field  and the metric can be defined in
 a general geometry, and Eq.(\ref{app2}) is generally true.

 The really new result that requires using the Einstein variational
 principle is the reduction of the tensor $A^i_{jk}$ to
 one vector $a_k$ (in fact, $A_k = (2\alpha + \beta) a_k$).
 The expression for $s_{ij}$ generalizing r.h.s. of (\ref{11b})
 can now be derived from the general equation:
 \be
 \label{app3}
 s_{jk} = \half (\nabla_j A_k + \nabla_k A_j) -
 \nabla_i A^i_{jk} + A^r_{ij} A^i_{rk} - A^r_{jk} A_r \,.
 \ee
 The terms linear in $A$ are equal to
 \be
 \label{app4}
 \half \bigg[(\alpha + \beta) (\nabla_j a_k + \nabla_k a_j) +
 (\alpha - 2\beta) \nabla_i a^i \biggl] \,,
 \ee
 and the quadratic terms are
 \be
 \label{app5}
 \quart \bigg[a_i a_k \,[(\alpha - 2\beta)^2 -3\alpha^2] \,+\,
 2 \,g_{ik} a^2 (\alpha - 2\beta) (\alpha + \beta)\biggl] \,.
 \ee
 When $\alpha = -\beta = -\third$, we reproduce Einstein's expression
 for $s_{jk}$.
 One can see that the sign of the first term in (\ref{app5})
 may be positive or negative but, in general, the second term in (\ref{app5})
 and  the linear terms in (\ref{app3}) do  not vanish.
 Apparently, one could get models with the positive sign
 of $\mu^2$ but this requires
 a more careful consideration which will be published elsewhere.

\bigskip
\bigskip

 {\bf Acknowledgment:}
 It is a pleasure for the author to thank for useful remarks
 V.~de~Alfaro, A.~Linde, V.~Rubakov, A.~Starobinsky,
 S.~Vernov and E.~Witten. The kind hospitality of
 M.~Mueller-Preussker and
  J.~Plefka at the Humboldt University (Berlin) is cordially appreciated.

 This work was also supported in part by the Russian Foundation for Basic
 Research (Grant No. 09-02-12417 ofi-M).

%\bigskip
%\bigskip
%\newpage

\end{document}